\documentclass[10pt, conference, compsocconf]{IEEEtran}
\usepackage{graphicx, url, fancyvrb, subfigure}

\usepackage{hyperref}

\ifCLASSINFOpdf
\else
\fi

\usepackage{algpseudocode}

\usepackage{xspace}
\newcommand{\sysname}{CloudForecast\xspace}

\begin{document}

%
\title{Location, Location, Location: Data-Intensive Distributed Computing in the Cloud}


\author{\IEEEauthorblockN{Michael Luckeneder and Adam Barker}
\IEEEauthorblockA{School of Computer Science\\
University of St Andrews\\
St Andrews, United Kingdom\\
Email: adam.barker@st-andrews.ac.uk}
}


%


\maketitle

\begin{abstract}

When orchestrating highly distributed and data-intensive Web service workflows the geographical placement of the orchestration engine can greatly affect the overall performance of a workflow. Orchestration engines are typically run from within an organisations' network, and may have to transfer data across long geographical distances, which in turn increases execution time and degrades the overall performance of a workflow. In this paper we present \sysname: a Web service framework and analysis tool which given a workflow specification, computes the optimal Amazon EC2 Cloud region to automatically deploy the orchestration engine and execute the workflow.  We use geographical distance of the workflow, network latency and HTTP round-trip time between Amazon Cloud regions and the workflow nodes to find a ranking of Cloud regions. This combined set of simple metrics effectively predicts where the workflow orchestration engine should be deployed in order to reduce overall execution time.

We evaluate our approach by executing randomly generated data-intensive workflows deployed on the PlanetLab platform in order to rank Amazon EC2 Cloud regions. Our experimental results show that our proposed optimisation strategy, depending on the particular workflow, can speed up execution time on average by 82.25\% compared to local execution. We also show that the standard deviation of execution time is reduced by an average of almost 65\% using the optimisation strategy.

\end{abstract}


%
\IEEEpeerreviewmaketitle

\section{Introduction}
Scientific workflows \cite{ppam08} are typically orchestrated using a workflow engine\footnote{e.g. \url{http://www.taverna.org.uk}} running locally within an organisation's network. However, if the Web services in the workflow are data-intensive and spread across many geographical regions, the data might have to move long distances in order to flow from the data sources to the Web services via the orchestrator. This in turn slows down the execution of the workflow and degrades the overall performance.

A possible solution to this problem could be to ``move'' the workflow orchestrator closer to the data and the Web services. Cloud environments provide a cost-effective platform for scientists and engineers to execute their workflows in a remote data centre as demonstrated by recent research \cite{deelman_EC2, deelman_workflow, trident_cloud, cost_cloud, cloud_hpc}. Using an Infrastructure as a Service (IaaS) Cloud, such as an Amazon EC2 instance, it is possible to automatically deploy the orchestrator into a suitable EC2 region that is ``closer'' to the data source and Web service nodes; in turn data would not have to travel as far and therefore execution times could be reduced.

The interesting problem arises when workflows consist of a large number of Web service nodes - all of them in different geographical regions. In these cases it is very challenging to judge where the closest and thus best-performing Cloud region might be. Furthermore, a certain Cloud region might be geographically closer to the Web service nodes but, due to a high network latency on a certain network link, using another Cloud region would actually result in a lower execution time.

In this paper we design, implement and evaluate \sysname: a pre-deployment analysis tool which can dynamically compute the ``optimal'' Cloud region to deploy the orchestrator, given a specific workflow consisting of multiple distributed services. We focus primarily on Directed Acyclic Graphs (DAG) based workflows since these are heavily used in the scientific community \cite{cite3}. DAGs present a dataflow view where data are the primarily concern, workflows are constructed from data processing (vertices) and data transport (edges). We consider different factors which could potentially affect the suitability of choosing a certain Cloud region: total geographical distance of workflow, network latency and HTTP round-trip time.

We develop methods to:
\begin{itemize}
  \item automatically deploy workflows given the topology of the workflow and a number of fixed resource locations.
  \item automatically evaluate and compare the performance of different workflow orchestration approaches.
  \item evaluate different factors which influence workflow execution time.
\end{itemize}

The remainder of this paper is structured as follows: in Section \ref{sec:preanalysis} we will outline the theoretical architecture of an analysis tool and a suitable workflow specification model followed by a brief discussion about the implementation. We then explain in Section \ref{sec:performanceanalysis} how we built a testing framework using PlanetLab and Amazon EC2 to evaluate the approach and the \sysname tool. Finally, in Section \ref{sec:results}, we discuss the results of the experimentation and how \sysname significantly reduced execution times.

\section{Pre-Deployment Analysis}
\label{sec:preanalysis}
In this section we outline the abstract architecture and a workflow specification model. We then discuss the \sysname tool and a worked example demonstrating how the optimisation is used in practice.

\subsection{Architecture}

\subsubsection{Workflow Specification}
In order to correctly analyse the workflow and find the optimised Cloud region to deploy the orchestration engine to, a workflow specification model is required. This should specify data sources, intermediate processing steps as well as a data sink. Since scientific workflows are inherently graph-based, the specification language should ideally also be graph-based so that it can easily be interpreted by \sysname as well as a workflow orchestrator. The workflow specification should also make it computationally simple to retrieve an ordered set of all distinct workflow nodes. This is an important feature required by the analysis engine.


\subsubsection{\sysname}

\begin{figure}[h]
\centering
\includegraphics[width=0.4\textwidth]{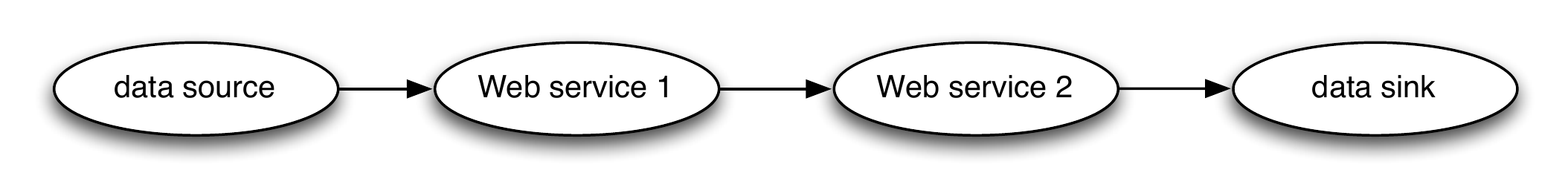}

\caption{Abstract workflow example}
\label{fig:abstract_example_workflow}
\end{figure}

\begin{figure}[h]
\centering
\includegraphics[width=0.4\textwidth]{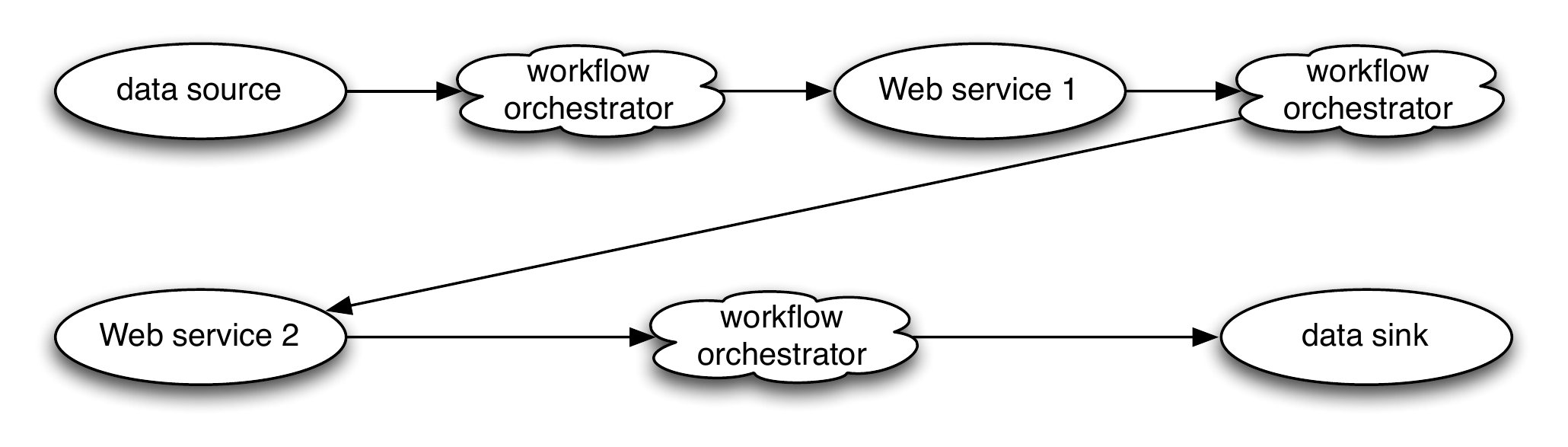}

\caption{Abstract candidate data flow graph}
\label{fig:abstract_candidate_wf_graph}
\end{figure}

The pre-deployment analysis tool takes a workflow specification and builds several candidate workflow graphs, which represent the data flow. The tool further requires a list of Cloud regions (e.g. Amazon EC2 regions) that should be considered for workflow execution. Every candidate graph will be based on a different one of these Cloud regions and represent a possible optimised workflow using the Cloud instance as the workflow orchestrator. Every edge represents the data flowing from a Web service node to the Cloud region or vice versa. Figure \ref{fig:abstract_candidate_wf_graph} illustrates a candidate graph built for the abstract workflow described in Figure \ref{fig:abstract_example_workflow}.

\sysname then gathers certain metrics for every graph in order to determine the optimal Cloud region. A metric value is required for every edge of the analysis tool's graph. Potential candidates for useful metrics include geographical distance, network latency (as measured by the UNIX ``ping'' command) and HTTP round-trip time (as measured by the UNIX ``curl'' command) between every pair of adjacent nodes.

Based on every candidate graph and metric used, \sysname computes an overall score. The final output is a separate table for every metric which ranks the Cloud regions by their predicted execution times. These tables are purely ordinal and no actual executions times are predicted. The pseudocode in Figure~\ref{eq:score} illustrates the algorithm. Note that the geographical distance is used for a preliminary ranking and then the top $n$ ranked regions are evaluated using ping and HTTP RTT.

\begin{figure}
  \begin{algorithmic}
    \Procedure{rank}{$nodes, aws\_regions, n=3$}
      \State $dag\gets build\_dag(nodes)$
      \State $cand\_graphs[~]\gets build\_cand(dag, aws\_regions)$
      \State $dist[~]\gets geo\_dist\_of\_each\_graph(cand\_graphs)$
      \State $top\_graphs\gets top\_n(cand\_graphs, dist, n)$\\

      \State $scores\gets [~]$
      \For{$g \in top\_graphs$}
        \State $scores[g] = \frac{total\_rtt(g) + total\_ping(g)}{2}$
      \EndFor
      \\
      \State \textbf{return} $scores, min(scores)$
    \EndProcedure
  \end{algorithmic}
\caption{Ranking Algorithm. $nodes$ is a list of web service node, $aws\_nodes$ is the list of distinct AWS regions and $n$ is the number of nodes to be used from the preliminary ranking (i.e. when $n=3$ the top 3 geographically ranked EC2 regions are used for the final calculation.}\label{eq:score}
\end{figure}

Finally, the Cloud region with the lowest overall score should be the best region to deploy the workflow orchestrator in order to minimise workflow execution time.

\subsection{Realisation}



\sysname is written in Python and is readily available as an open source project hosted on GitHub\footnote{\url{http://github.com/bigdatalab/movingdata}}. The code heavily uses the ``boto''\footnote{\url{http://github.com/boto/boto}} library to interface with the Amazon EC2 APIs. To build the internal representation of the workflow graph as well as the candidate graphs for different Cloud regions, we implemented a Directed Acyclic Graph (DAG) which can be traversed to calculate the sum of all edge weights.

In the actual implementation for the experiments the workflows are defined using concrete implementations of an abstract Python class rather than a graph-based markup language. This has the advantage that the workflow can be specified using normal Python commands and one is not restricted to using a more rigid, graph-based specification language.

The core of the tool is an abstraction of one to many Amazon EC2 instances in different Cloud regions. This means that with a class instantiation and a single method call, instances in multiple EC2 regions can be launched, SSH commands can be run, and the return values can be displayed simultaneously. Similarly, files can be transferred via SCP into multiple instances at once.

When the \sysname tool is executed given a certain workflow, the tool goes through three stages: pre-analysis, metric gathering and analysis.

In the \emph{pre-analysis stage}, it first uses the DAG implementation to create a series of graphs - one for each Amazon EC2 region and metric (24 graphs\footnote{3 metrics $\times$ 8 regions $=$ 24 graphs}). These graphs are implementations of the graph presented in Figure \ref{fig:abstract_candidate_wf_graph}. In these graphs, the edges correspond to the metrics that have to be gathered. Given this information, the tool enters the metric gathering stage.

The \emph{metric gathering stage} starts with launching EC2 instances in every Cloud region and deploying the metric script. For every edge in every graph, the metrics are retrieved: network latency is obtained using ``ping'' and HTTP RTT using ``curl''. To get geographical distances between nodes, the ipinfodb\footnote{\url{http://ipinfodb.com}} API and some spherical geometry calculations are used.

In the \emph{analysis stage}, \sysname traverses all the generated graphs and sums the edge weights. This will generate the Cloud region ranking tables for all three metrics. Furthermore, the final score is calculated by averaging the network latency and HTTP RTT metric for every Cloud region (see Figure \ref{eq:score}). These tables are then displayed, all Amazon EC2 instances previously launched are terminated and the analysis tool quits.

\subsection{Worked Example}
Here we cover a worked example of how \sysname generates a deployment decision for a simple workflow.



\begin{figure}[!t]
\centering
\includegraphics[width=2.2in]{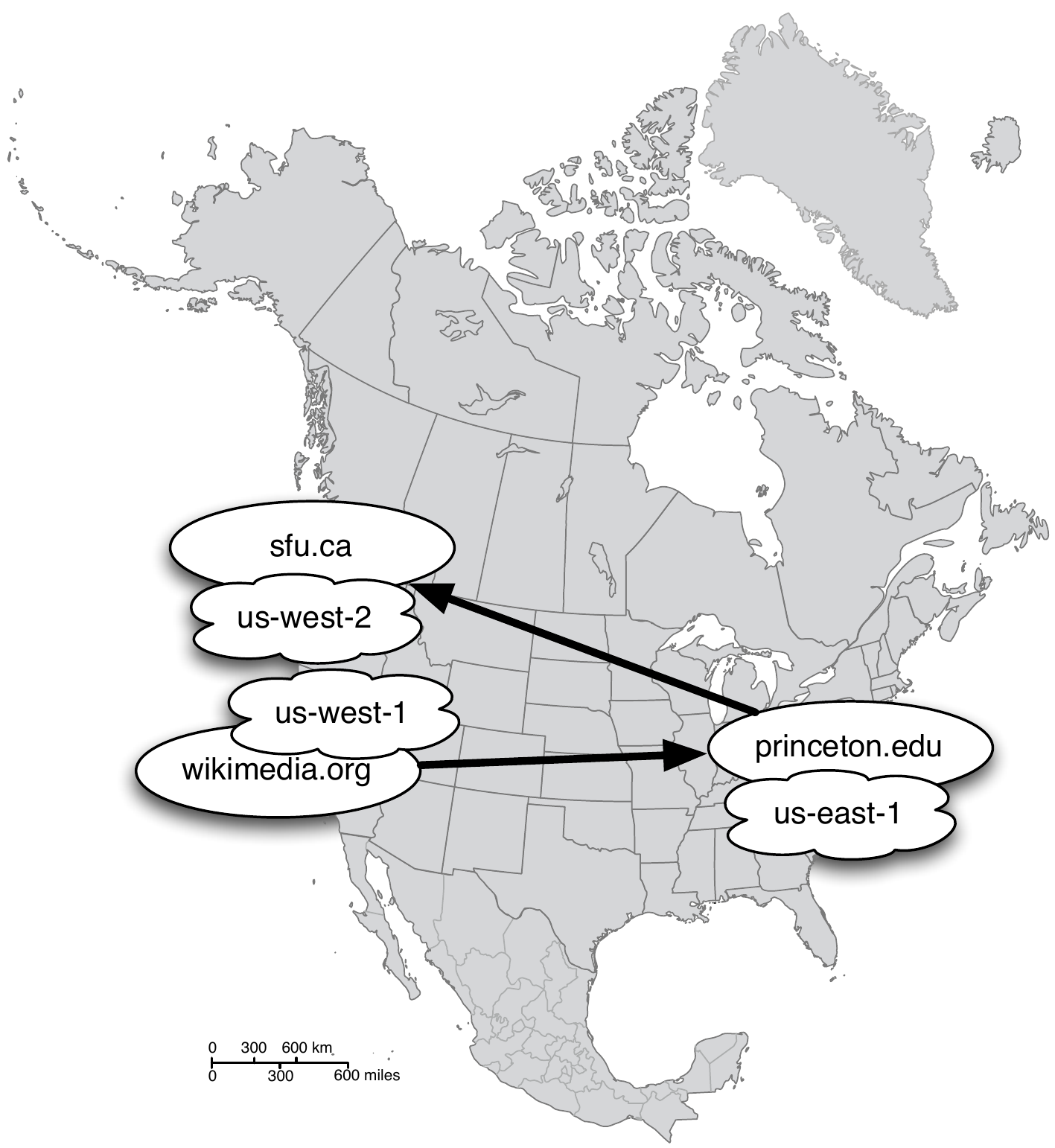}

\caption{Example workflow with closest Amazon EC2 regions}
\label{fig:example_workflow_with_ec2}
\end{figure}

Figure \ref{fig:example_workflow_with_ec2} shows a simple, sequential workflow with a data source (wikimedia.org) and two workflow nodes hosted on PlanetLab (planetlab-03.cs.princeton.edu, cs-planetlab4.cs.surrey.sfu.ca). It takes an image from wikimedia and then sends it to the princeton.edu node for processing. The result from this step is then sent to the sfu.ca node. The workflow is specified using plain Python by providing a concrete implementation of the abstract workflow specification class.

\begin{figure}[!t]
\centering
\includegraphics[width=0.4\textwidth]{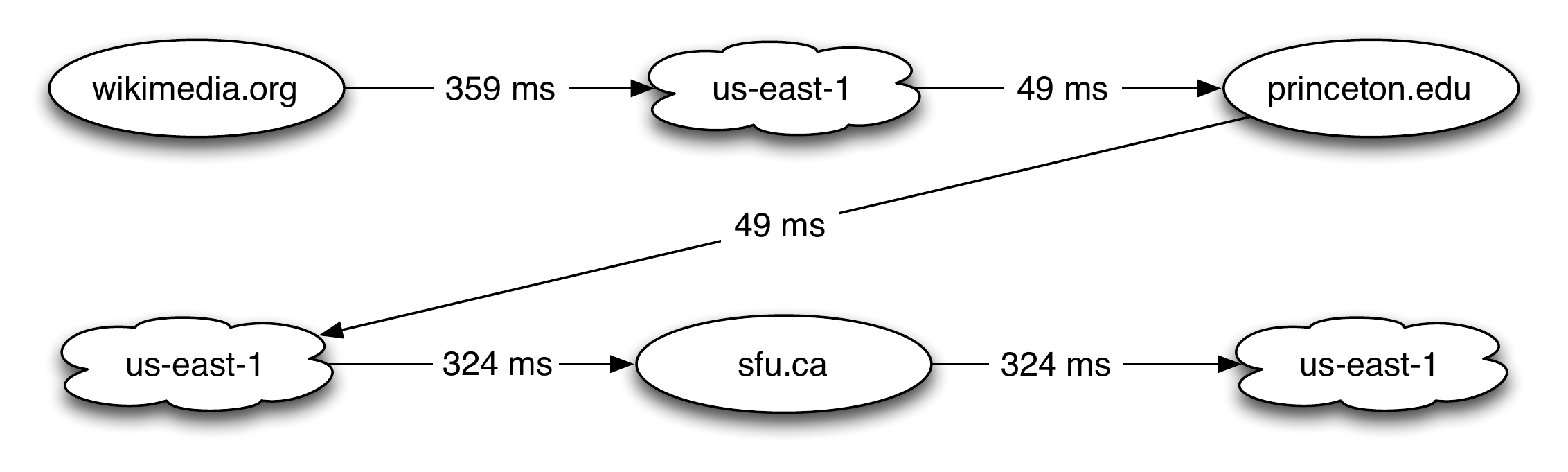}

\caption{Candidate workflow graph using EC2 region us-east-1 with network latency metric}
\label{fig:labeled_candidate_graph}
\end{figure}

In addition, Figure \ref{fig:example_workflow_with_ec2} also shows the workflow in relation to the closest Amazon EC2 Cloud regions (us-east-1, us-west-1, us-west-2). This illustrates the dilemma faced when deciding on the correct Cloud region to deploy the workflow orchestrator to: which Cloud region will result in the lowest execution time?

\sysname will take this workflow and build the internal candidate graphs for every metric and Cloud region. After all the metrics tools have terminated, the tool will have essentially labeled all edges in the graphs. Figure \ref{fig:labeled_candidate_graph} is an example of what those internal graphs look like.

The tool then evaluates the graphs and ranks the Cloud regions in order of increasing execution time. Table \ref{fig:example_result_table} shows the output of \sysname, an average of HTTP round-trip time and network latency (see Figure \ref{eq:score}).

Based on the final score rankings, we choose to deploy the workflow orchestrator in Cloud region us-east-1.


\begin{table}[h]
\renewcommand{\arraystretch}{1.2}
\caption{\sysname result output}
\label{fig:example_result_table}
\centering
\begin{tabular}{|l|l|}
\hline
EC2 endpoint & final score\\
\hline
us-east-1 & 92530.42\\
us-west-2 & 186251.487\\
us-west-1 & 186374.351\\
sa-east-1 & 366450.152\\
ap-northeast-1 & 421102.237\\
ap-northeast-2 & 510982.726 \\
ap-southeast-1 & 532180.129\\
eu-west-1 &500178094.532\\
\hline
\end{tabular}
\end{table}

\section{Performance Analysis}
\label{sec:performanceanalysis}
In this section we describe how we analysed the performance and correctness of our approach.

\subsection{Experimental Setup}
To verify the functionality of the \sysname tool, we randomly generated and analysed a set of DAG workflows; both simple sequential workflows and more complex parallel workflows involving multi-source and multi-sink configurations. Randomly generated workflows are realistic because usually scientists have no choice over the ordering of third-party services as they must be used in a certain order to execute an end-to-end distributed application. For a complete description and graphical representation of the workflows used in our experiments please refer to our homepage\footnote{\url{http://bigdata.cs.st-andrews.ac.uk/index.php/projects/\#cloudforecast}}.

In order to create a highly reusable and controlled Web service workflow, we decided to host a simple, sequential image processing workflow on the PlanetLab \cite{planetlab} framework (a global research network) and other servers. The idea is that these services can be invoked by a simple HTTP request with an image in the HTTP request. The Web service then executes some time-consuming image processing (rotating it) and then returns the image in the HTTP response.

In a typical workflow used in the experiments (e.g. Figure \ref{fig:example_workflow_with_ec2}), an image is downloaded from the wikimedia servers and then passed on to the first workflow node by the orchestrator. The processed image is then downloaded from the first node and sent to the second node where the same processing step is applied, and so on.

\subsubsection{PlanetLab (Test Web Services)}

In PlanetLab we used a total of 6 nodes, 5 of which were in North America and 1 of which was in continental Europe\footnote{PL nodes: Carnegie Mellon University, USA; Kansas State University, USA; Princeton University, USA; Simon Fraser University, Canada; University of Ljubljana, Slovenia; Williams College, USA}. These specific nodes were chosen as they offered the most reliable availability and performance to conduct the experiments. All nodes run a Python server script which listens for HTTP requests. On incoming HTTP requests with an image sent in the request, the script saves the image to the filesystem, rotates the image and then returns it in the HTTP response.



\subsubsection{Amazon EC2 (Workflow orchestration and analysis)}

For both the \sysname tool and the actual workflow orchestrator deployment, we used t1.micro instances running Ubuntu 12.04.1 LTS in the 8 different regions\footnote{EC2 Regions: North America: East \& West; South America; Europe; Asia Pacific: Sydney, Tokyo, Singapore}. The computation power and storage provided by the t1.micro instances was sufficient for the experiments as the \sysname tool and the workflow orchestrator to have very small memory footprints and small computation power requirements.

\subsubsection{Random Workflows}
The testing workflows were created by randomly selecting (with replacement) a sequence of nodes from a list of the available Web service nodes (PlanetLab and the other servers). These workflows are stored in plain-text files, with every line containing a single node. All workflows use wikimedia.org as their data source. 7 workflows were generated - with 2, 3, 4, 5, 10 and 12 nodes. In order to evaluate more complex workflows, such as multiple source patterns, two additional workflows with 7 and 13 nodes were generated and included in the test suite.

Due to the lack of consistently available PlanetLab nodes in South America, Africa or Asia, we focus on North America and Europe. We generated sample workflows with 2, 3, 4, 5, 7, 10, 12 and 13 nodes.

\subsubsection{The Verification Framework}

In order to verify the optimisations proposed by \sysname, we implemented a very simplified, bare-bones workflow orchestrator. It repeatedly executes the workflow a pre-defined number of times and returns the total execution time. The orchestrator can easily be executed locally as well as remotely on Amazon EC2 instances via SSH. This enables us to launch the workflow in the predicted Cloud region as well as from within the university network and verify if the Cloud optimised version is faster.

\section{Results}
\label{sec:results}
In this section we discuss the experimental results we obtained through evaluation the previously generated workflows using \sysname.

\subsection{General results}


\begin{table*}[!t]
\renewcommand{\arraystretch}{1.2}
\caption{Workflow execution results}
\label{tab:wf_execution}
\centering
\begin{tabular}{|ccc|||cc||cc|cc||cc|cc|}
\multicolumn{3}{c}{}& \multicolumn{2}{c}{local execution (s)} & \multicolumn{2}{c}{2nd ranked (s)} & \multicolumn{2}{c}{2nd ranked vs local (s)} & \multicolumn{2}{c}{1st ranked (s)} & \multicolumn{2}{c}{1st ranked vs local (s)} \\
\hline
workflow & nodes & data (MB) & mean & $\sigma$ & mean & $\sigma$ & speedup &  $\Delta\sigma$ & mean &  $\sigma$ & speedup & $\Delta\sigma$\\
\hline
A  & 2 &   74   &124.86 & 36.05 & 110.75 &   40.29 &  13\%  &  11.76\%  & 48.31    & 0.24   &   159\% & -99.33\%     \\
B  & 3 &   111   &66.10 & 16.45 & 87.81 &      3.77 &  -25\%  & -77.08\%  & 22.93    & 1.73   &  188\% &  89.48\%     \\
C  & 4 &   148   &118.85 & 14.82 & 178.56 &    7.70 &  -33\%  &  -48.04\% & 115.74  & 6.02   &   3\%   &  -59.38\%    \\
D  & 5 &   185   &339.14 & 64.73 & 288.05 &   25.32 &  18\%  &   -60.88\% & 276.13    & 19.75   &23\%  &   -69.49\%      \\
E  & 5 &   185   &515.30 & 154.85 & 400.45 &   39.70 &  29\%  &   -74.36\ & 356.71    & 14.90   &44\%  &  -90.38\%       \\
F  & 10 &  370   & 560.88 & 32.40 & 470.45 &   39.95 & 19\%   &  23.30\%  & 358.08    & 21.35   &57\%  &   -34.10\%     \\
G  & 12 &  444   & 631.00 & 15.34 & 585.62 &  30.85 &   8\%  &   101.11\% & 444.30   & 4.82   &  42\%  &  -68.58\%     \\
H & 7 & 435 & 340.95 & 45.60 &  301.57 & 55.44 & 13\% & 21.58\%       & 221.93 & 54.21 & 54\% & 18.88\% \\
I & 13 & 1220 & 1604.83 & 1029.28       &  635.04 & 39.18 & 153\% & -96.19\%     & 591.61 & 101.35 & 171\% & -90.15\%\\
\hline
MEAN & && 477.99 & 156.61 &        339.81 &  31.36 & 21.53\% & -22.09\%      &270.64 &  24.93 & 82.25\% & -64.67\%\\
\hline
\end{tabular}
\end{table*}

Table \ref{tab:wf_execution} summarises the different workflow execution times. Every workflow was executed 5 times, on different days and different times of day to avoid systematic errors. For local execution, the table presents mean and standard deviation of execution time. For the first and second-ranked regions the table shows mean and standard deviation of execution, mean percentage speedup compared to local execution and the percentage decrease in standard deviation compared to local execution.

\begin{figure*}[htp]
\begin{center}
\label{fig:boxplots}

\subfigure[Workflow A; 2 nodes]{\includegraphics[width=0.3\textwidth]{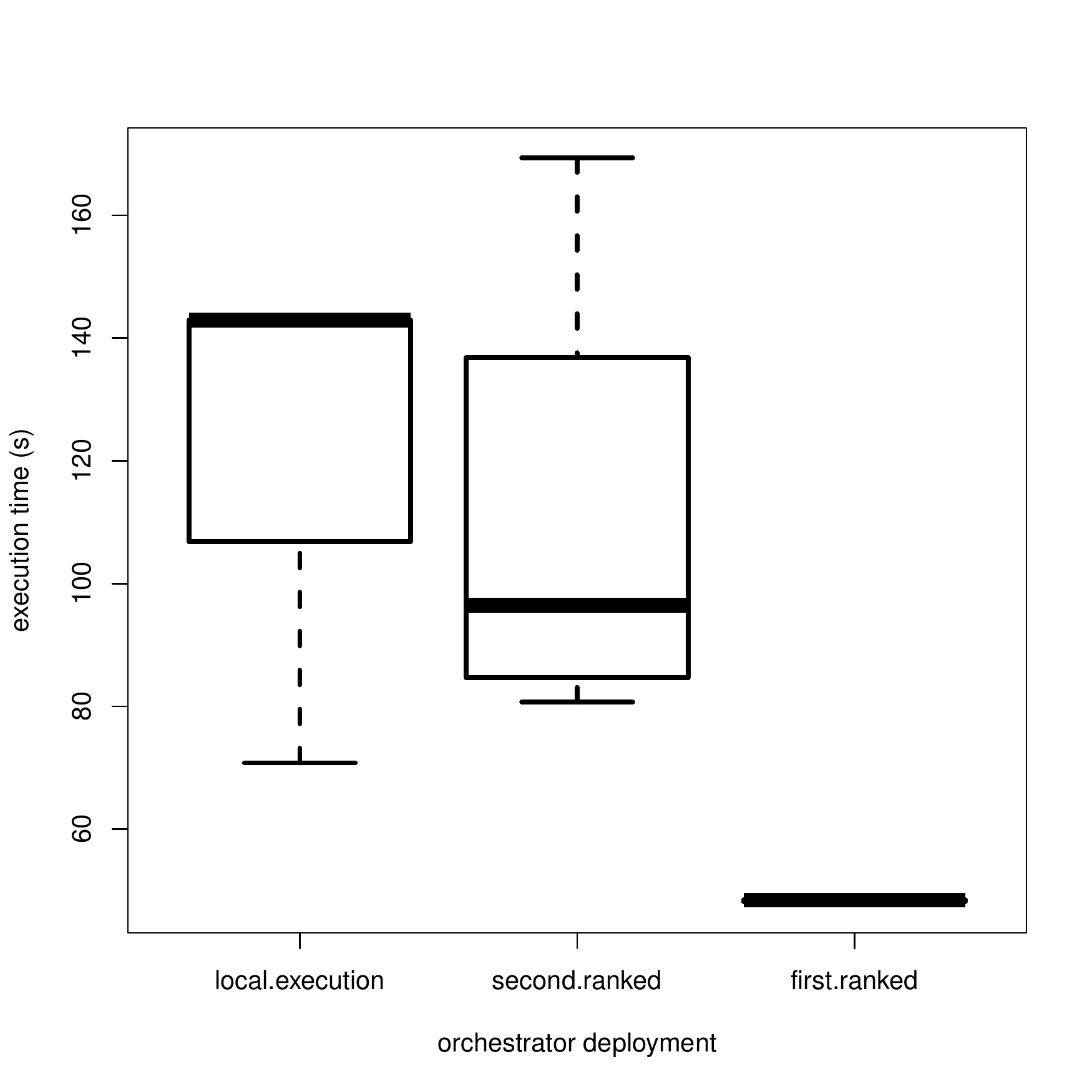} \label{fig:a1}}
\subfigure[Workflow B; 3 nodes]{\includegraphics[width=0.3\textwidth]{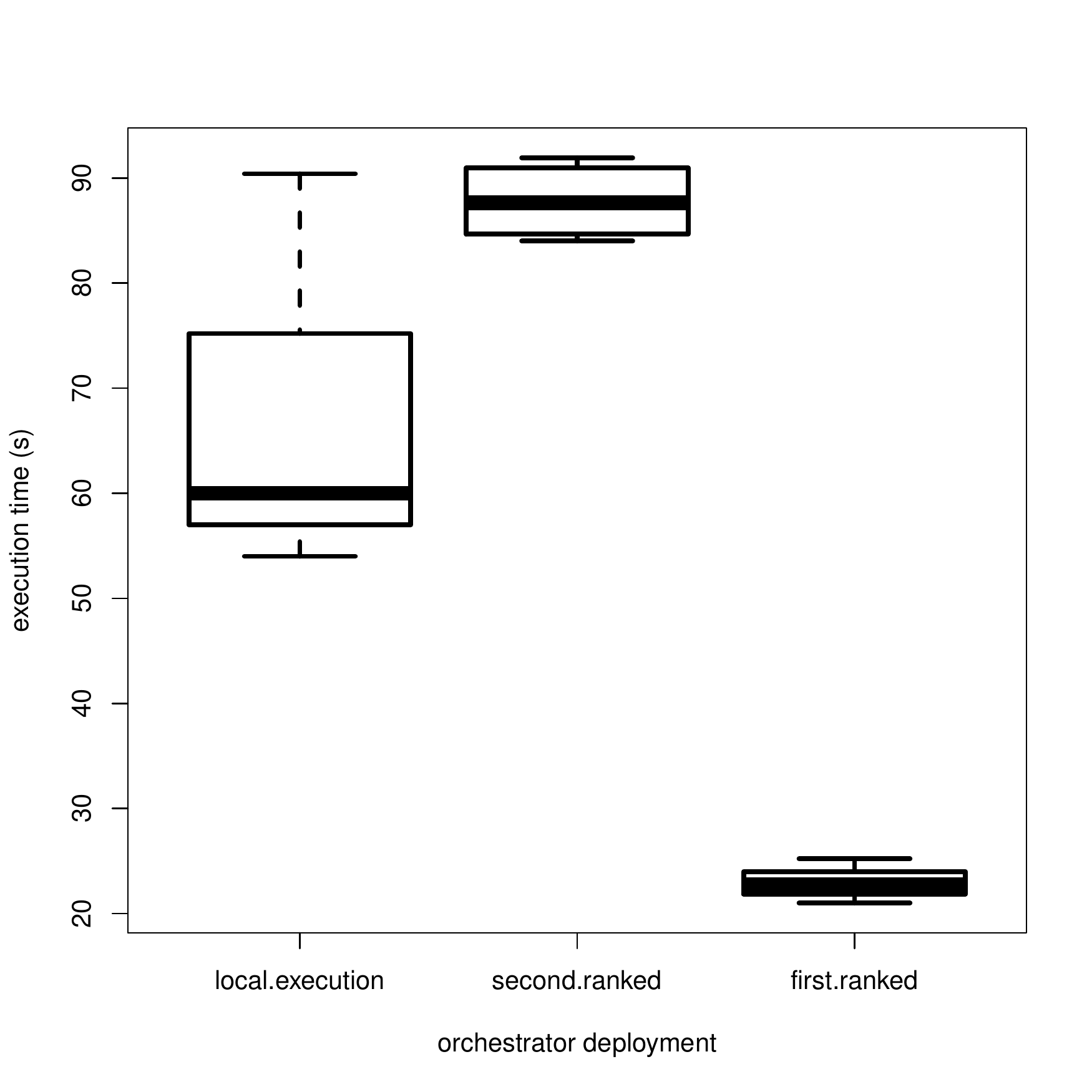} \label{fig:a2}}
\subfigure[Workflow C; 4 nodes]{\includegraphics[width=0.3\textwidth]{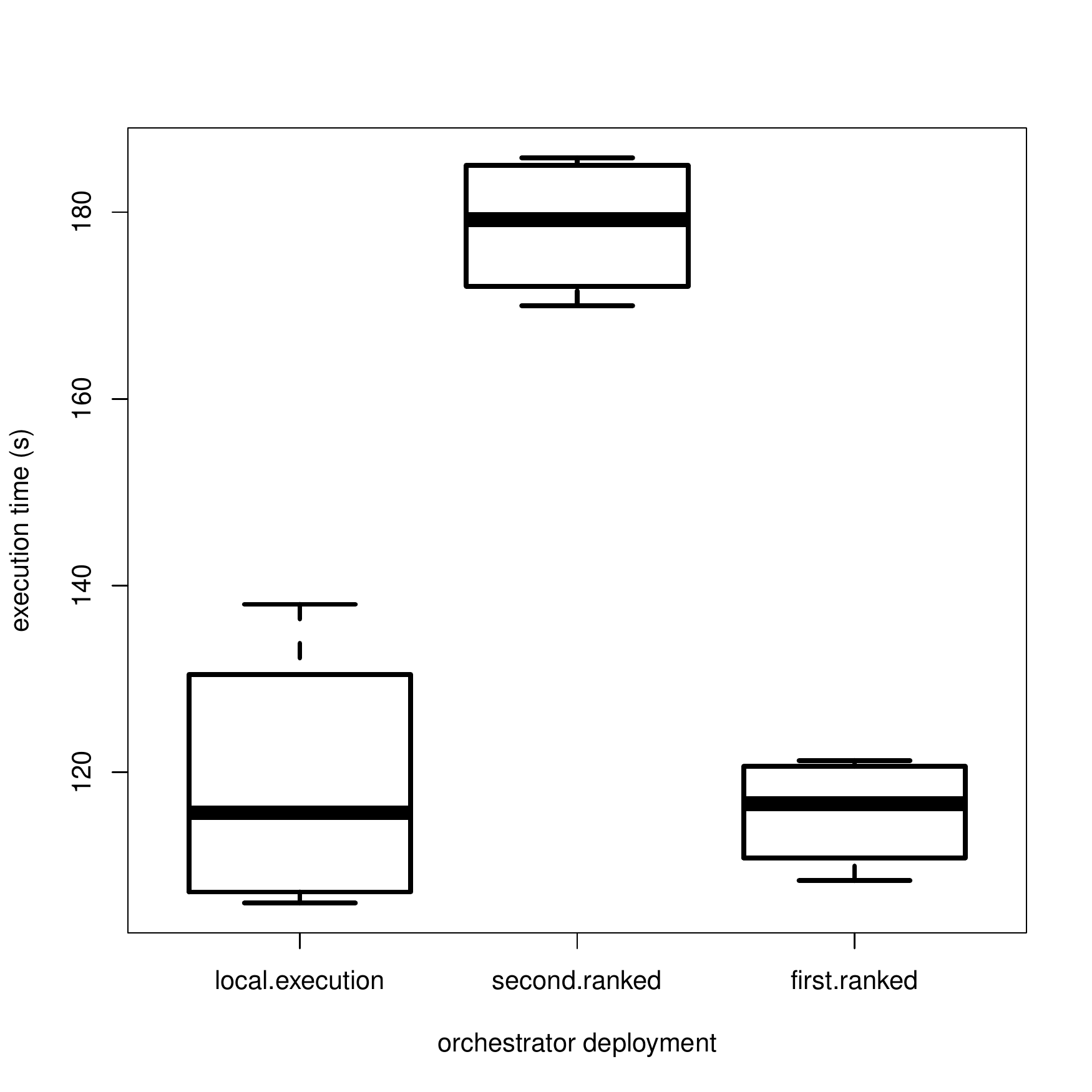} \label{fig:a3}}

\subfigure[Workflow D; 5 nodes]{\includegraphics[width=0.3\textwidth]{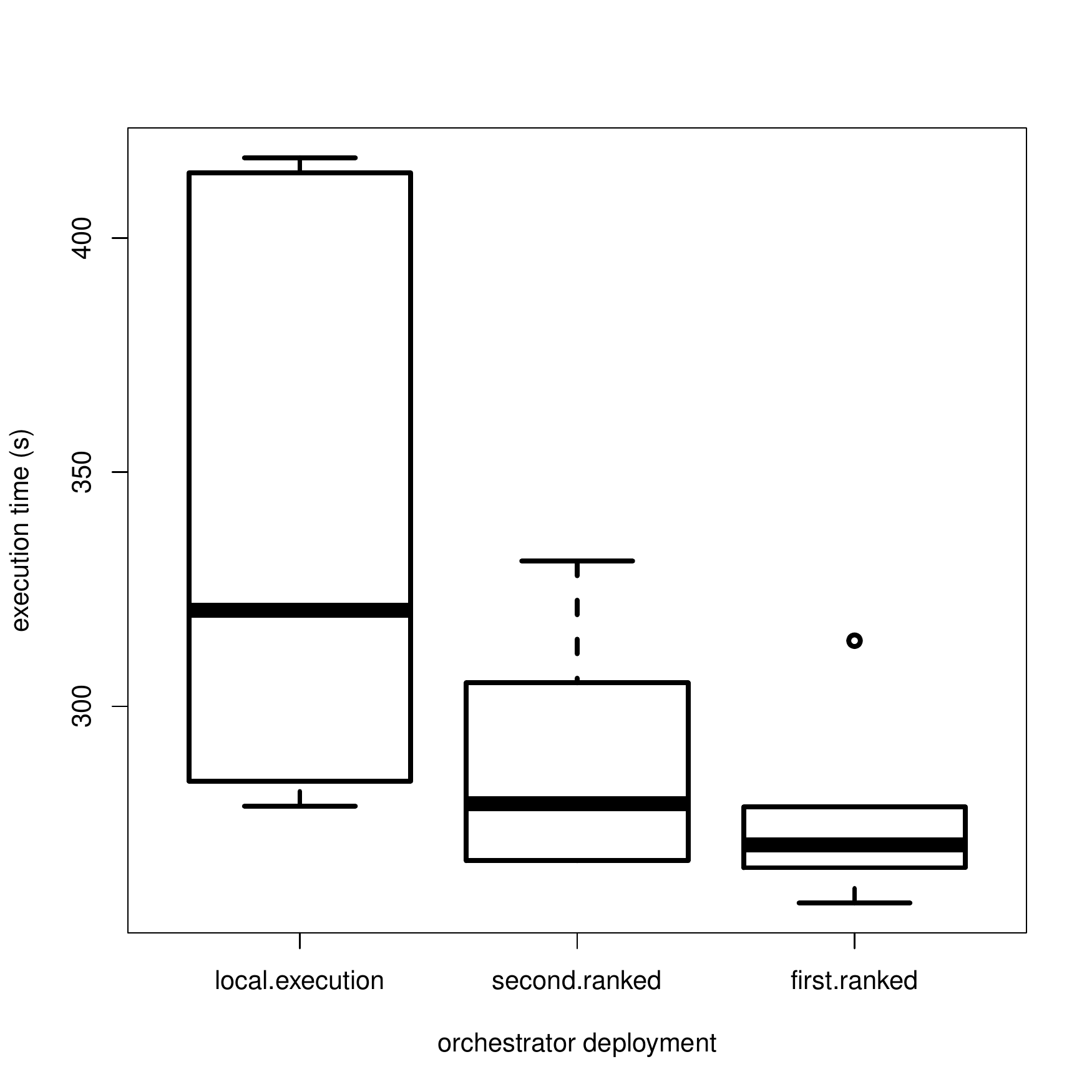} \label{fig:a4}}
\subfigure[Workflow E; 5 nodes]{\includegraphics[width=0.3\textwidth]{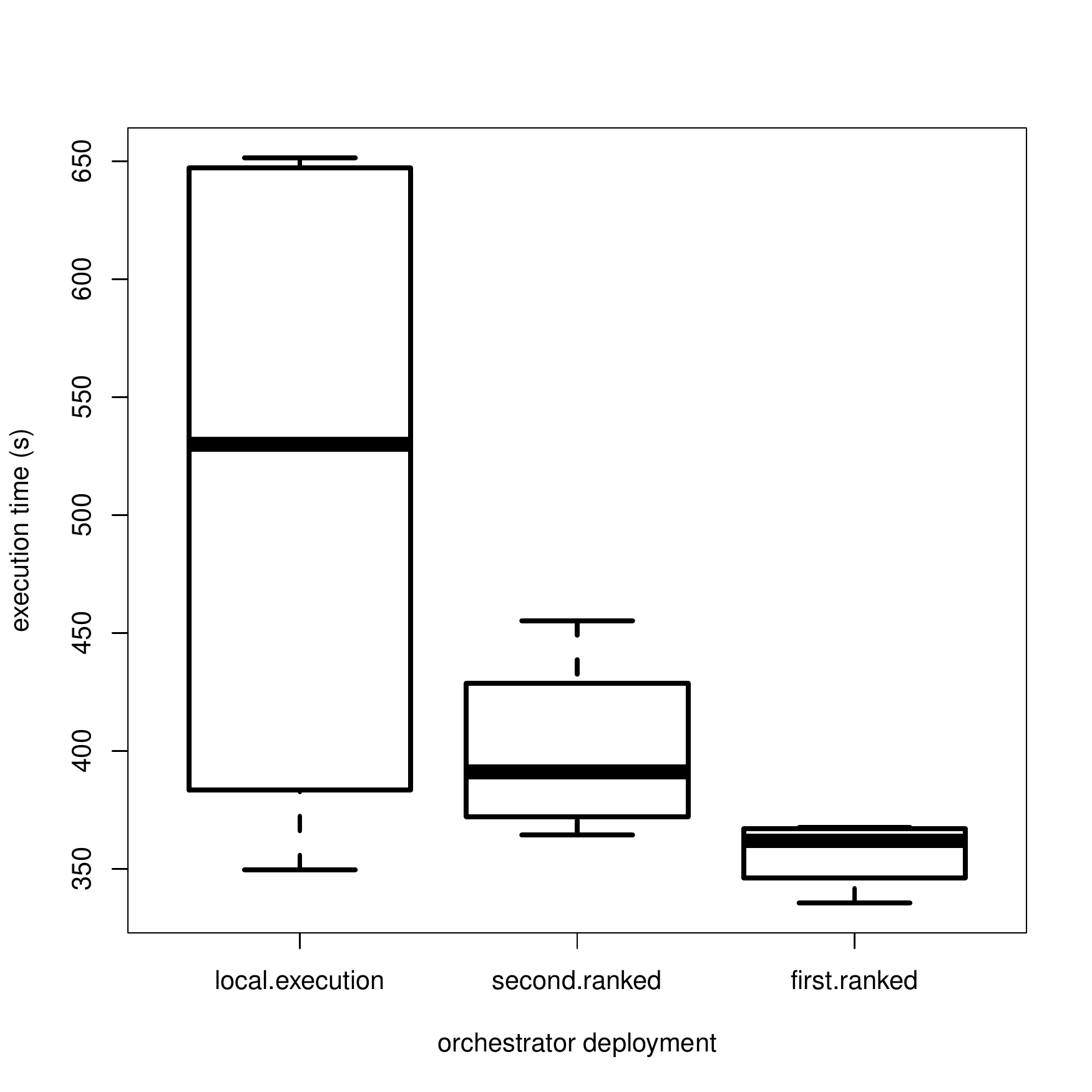} \label{fig:a5}}
\subfigure[Workflow F; 10 nodes]{\includegraphics[width=0.3\textwidth]{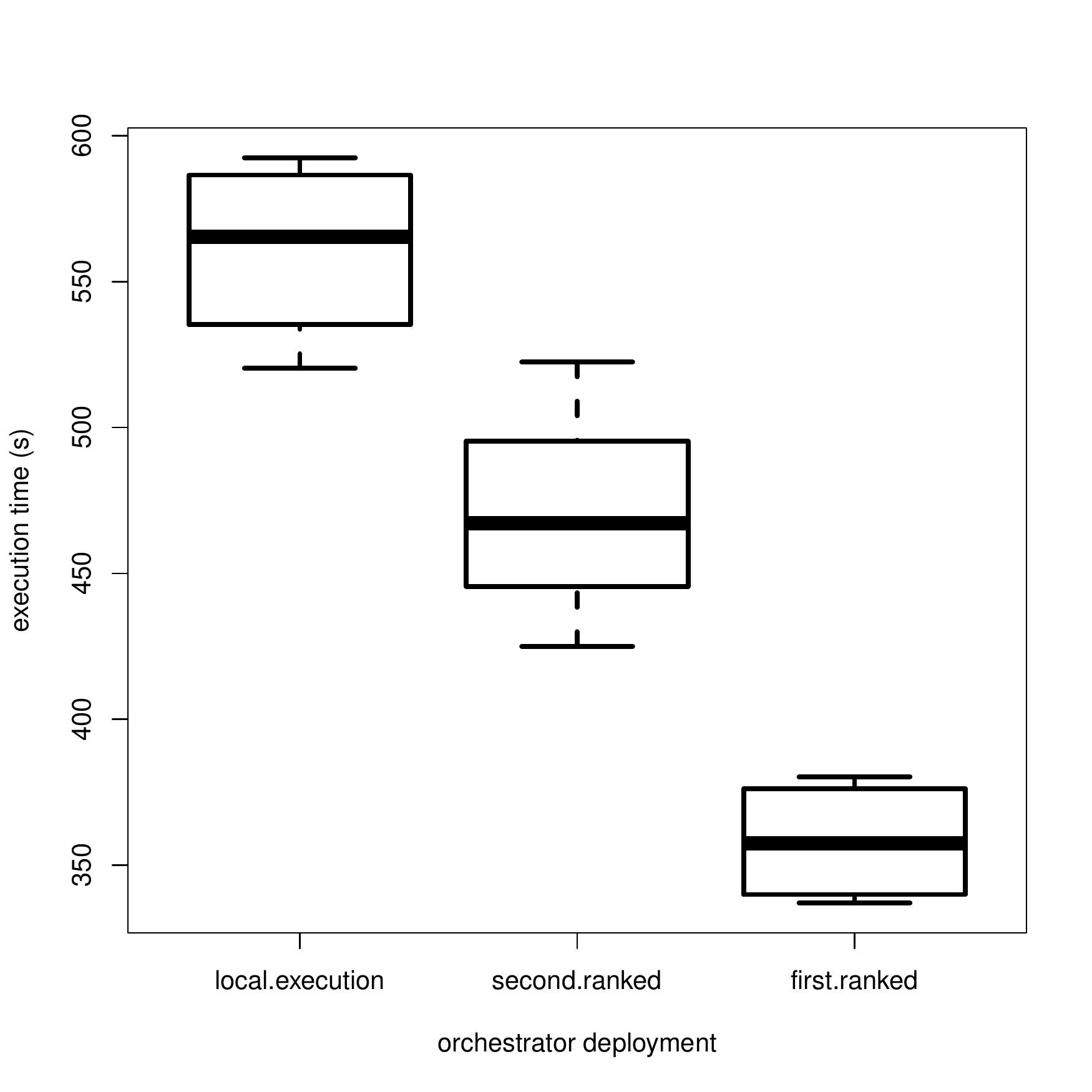} \label{fig:a6}}

\subfigure[Workflow G; 12 nodes]{\includegraphics[width=0.3\textwidth]{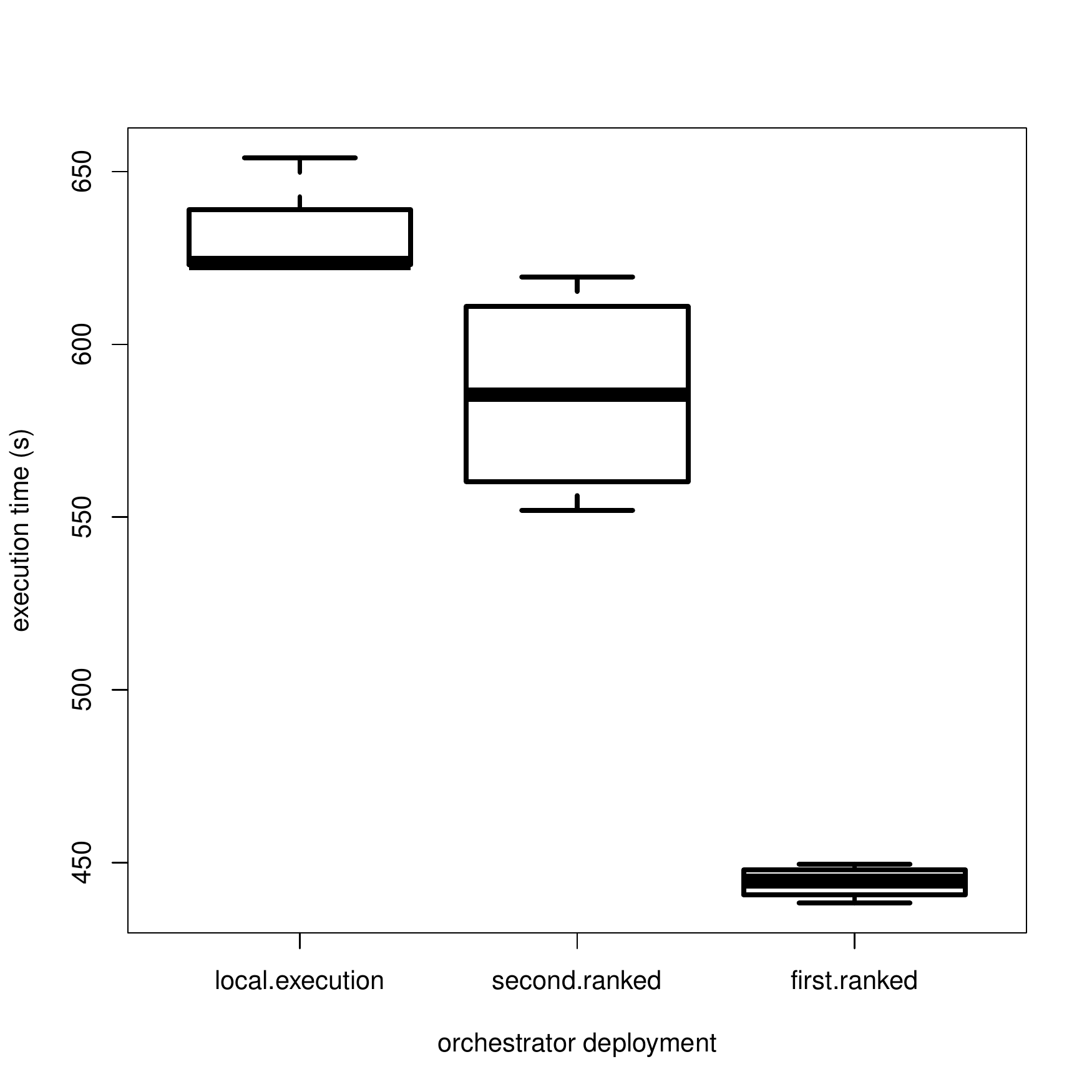} \label{fig:a7}}
\subfigure[Workflow H; 7 nodes]{\includegraphics[width=0.3\textwidth]{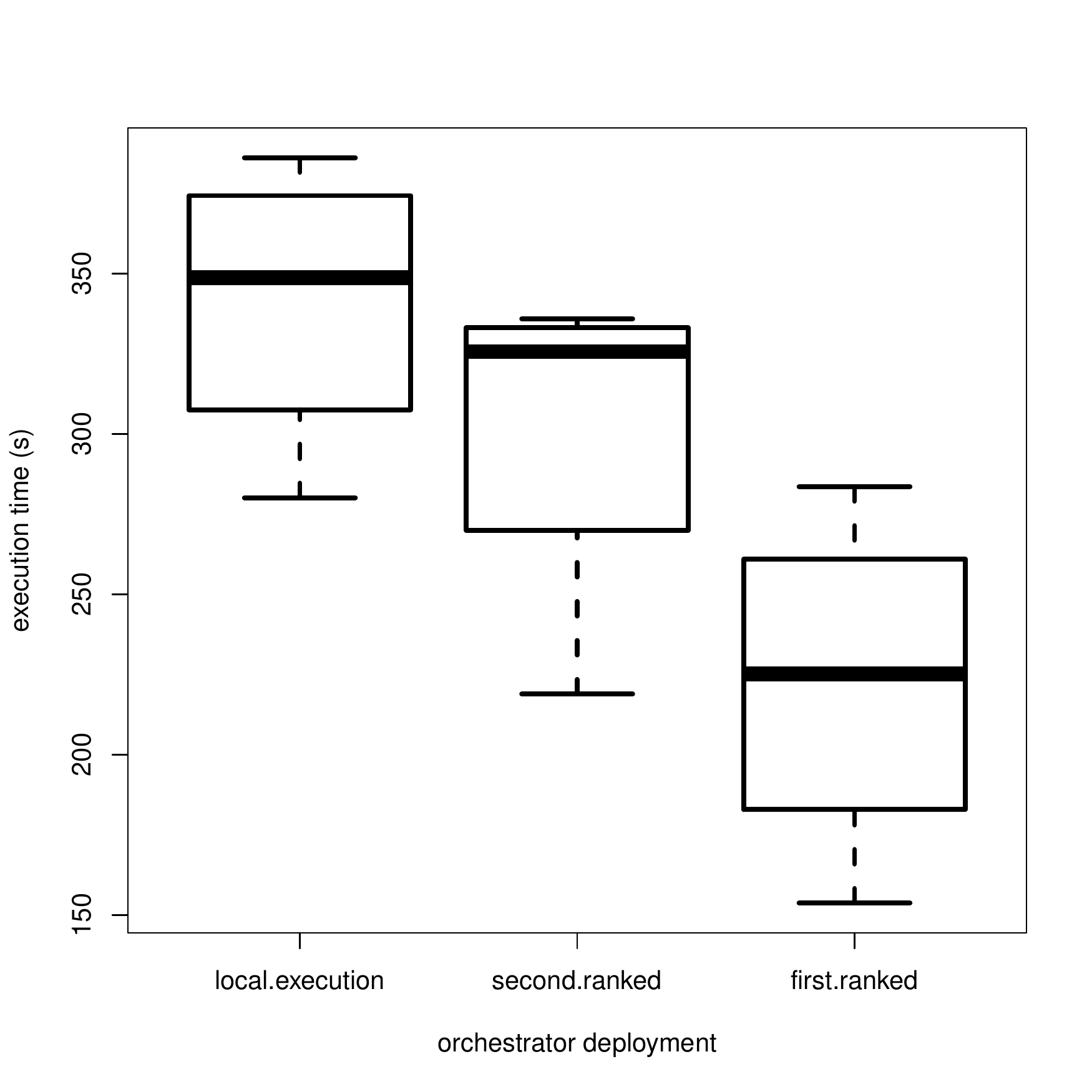} \label{fig:a8}}
\subfigure[Workflow I; 13 nodes]{\includegraphics[width=0.3\textwidth]{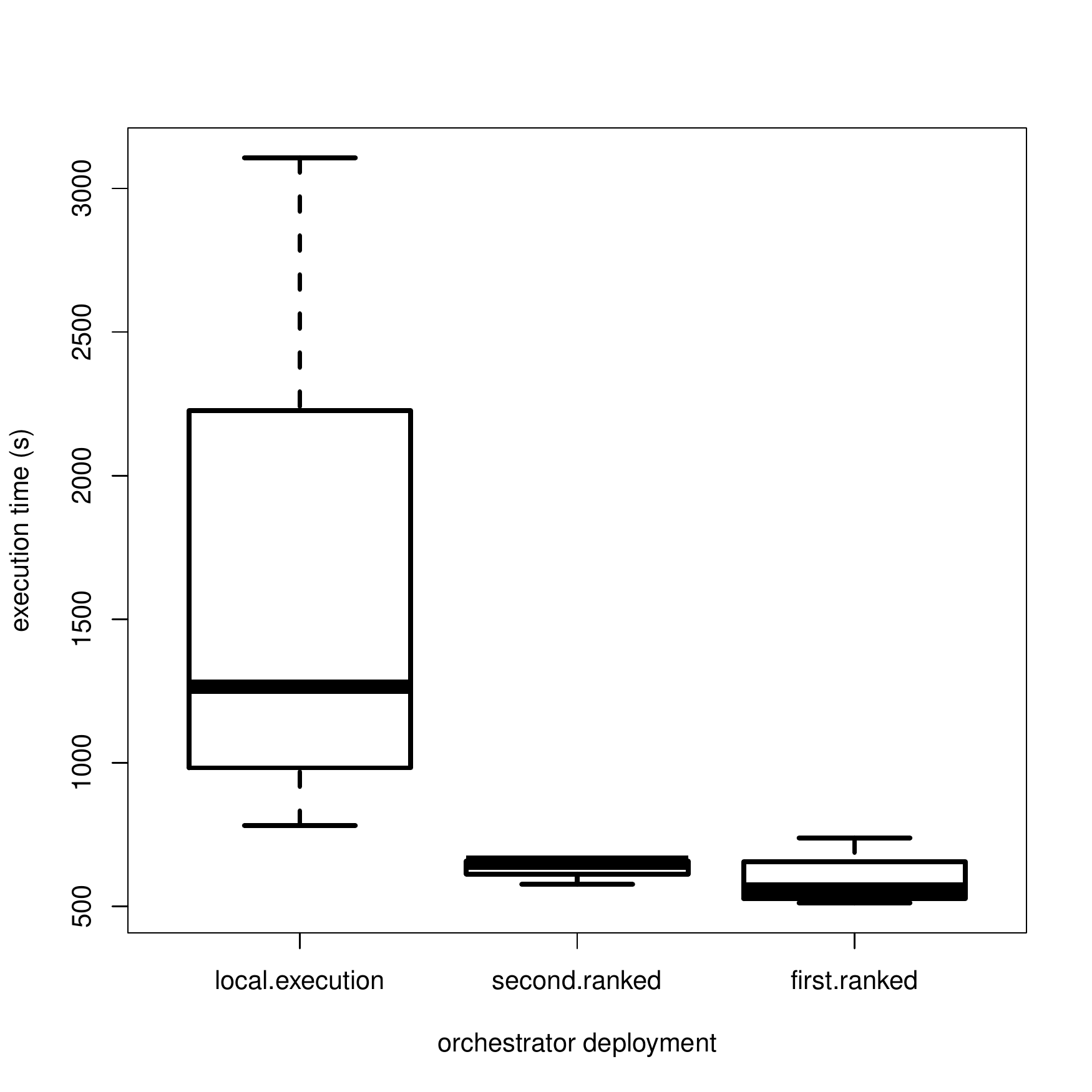} \label{fig:a9}}
\subfigure[Performance gain; nodes in brackets]{\includegraphics[width=0.3\textwidth]{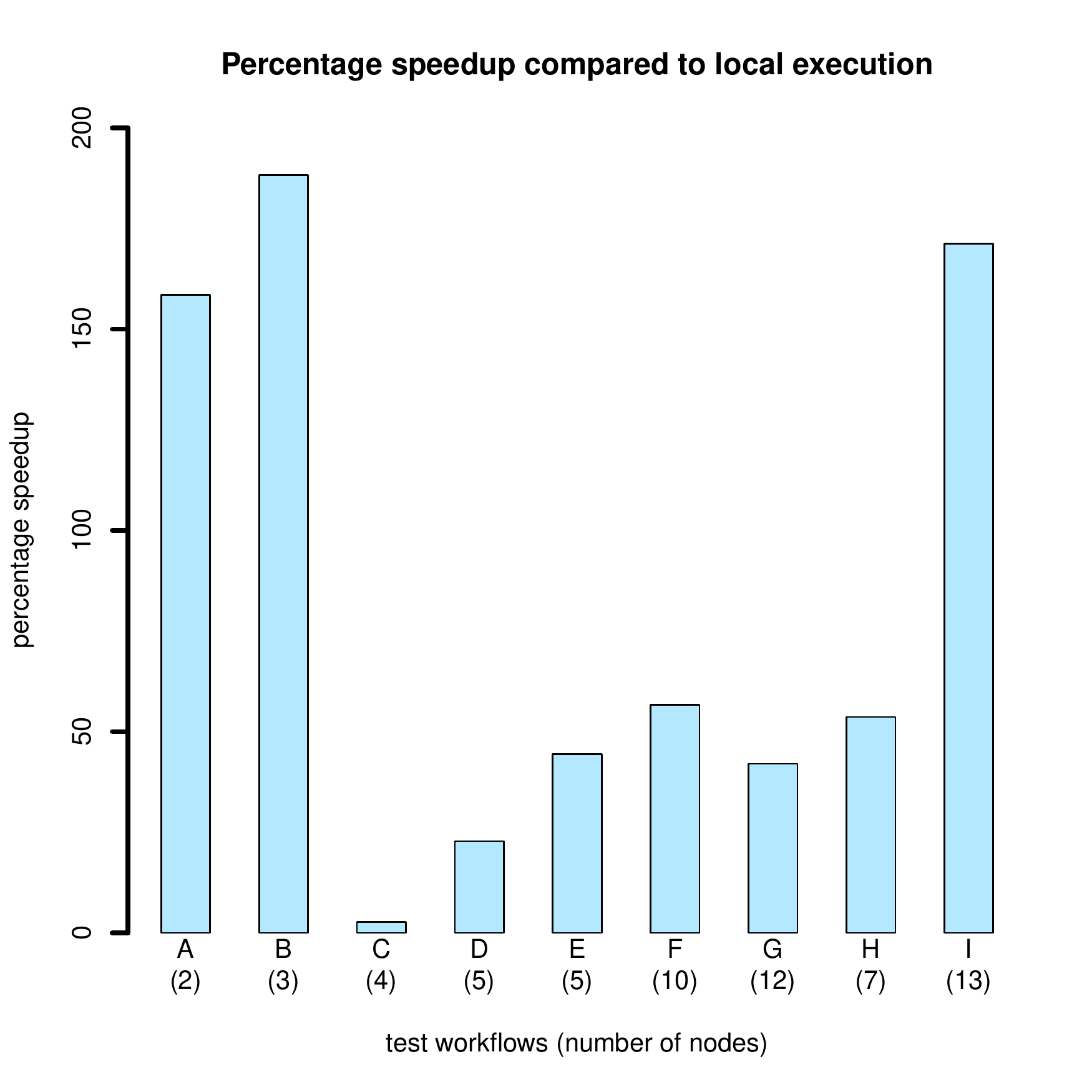} \label{fig:overall_performance_gain}}

\caption{Execution times of the sample workflows}
\end{center}
\end{figure*}

The box plots in Figure \ref{fig:a1} - \ref{fig:a9} graphically illustrate the results of Table \ref{tab:wf_execution}. We can clearly observe that the mean execution times are greatly reduced by orchestrating them in the highest-ranked Cloud region as opposed to executing them locally. However, the data also shows that the magnitude of reduction in execution time highly depends on the workflow being analysed. Especially Figure \ref{fig:a3} shows that for this particular workflow, local execution time is very close to the Cloud-optimised execution time.

Figure \ref{fig:overall_performance_gain} illustrates the speedup in mean execution time for each sample workflow due to being run in the first-ranked Cloud region compared to local execution. The speedups range from $3\%$ to $188\%$ with a mean of $82.25\%$. In contrast, when the workflow is deployed in the second-ranked Cloud region, the mean speedup from local execution is only $21.53\%$. We can conclude that the analysis correctly ranks the Cloud regions to reduce execution time.

We also note from Table \ref{tab:wf_execution} that the standard deviation of execution is reduced by an average of almost $65\%$ when run in the first-ranked region. This leads to the conclusion that the highest-ranked Cloud region as calculated by the \sysname tool makes execution times more stable compared to local execution.

\subsection{Factors}

Table \ref{tab:wf_execution} showed that the first-ranked Cloud regions were consistently faster than the second-ranked Cloud regions. Here, we discuss the significance of the individual metric factors used by the \sysname tool to rank the data centres.

Table \ref{tab:factor_predictions} summarises whether a specific metric correctly predicted the best performing Cloud region.

\begin{table}[!t]
\renewcommand{\arraystretch}{1.2}
\caption{Correct predictions}
\label{tab:factor_predictions}
\centering
\begin{tabular}{|c|c|c|c||c|}
\hline
workflow  & total distance & latency & HTTP RTT & overall\\
\hline
A   & yes & yes & no & yes     \\
B   & yes & yes & yes & yes    \\
C   & yes & yes & yes & yes     \\
D   & yes & yes & yes & yes     \\
E   & yes & yes & no & yes     \\
F   & yes & yes & yes & yes     \\
G   & yes & yes & yes & yes    \\
H   & yes & yes & yes & yes\\
I   & yes & yes & no  & yes\\
\hline
\end{tabular}
\end{table}

\subsubsection{Geographical distance}
Geographical distance seems to give a consistent estimate of the best Cloud region to deploy the workflow orchestrator. Based on the results of our experiments, we can conclude that total geographical distance of a workflow on its own is already a very good indicator to rank Cloud regions.

However, distance analysis is static and does not take into account unexpected network latencies on specific network links. Thus, geographical distance should only serve as a crude indicator to rank Cloud regions and we chose not to include this metric in the overall score calculation.

\subsubsection{Network latency}
Network latency, as measured by average ICMP ping times, also seems to be consistent in predicting the best performing Cloud region. However, especially when services are hosted on big server farms, the ping might only measure latency in the Internet and not in the network behind the gateway of the Web service.

\subsubsection{HTTP round-trip time}
Since network latency may not take into account the private network and application layer latencies, we included HTTP round-trip time as a potential factor to rank Cloud regions.
HTTP round-trip time, as measured by a single request to the endpoint URL using ``curl'', is useful to rank Cloud regions in some instances. There are two sample workflows, however, where the RTT prediction was incorrect. Therefore, this metric is only partially useful.

\subsubsection{Overall ranking}

Although relatively simple metrics have been used, this combined score obtained by averaging ping and RTT scores (see Figure \ref{eq:score}) is a consistent indicator of Cloud region performance for the specific sample workflow. Throughout our experimentation we could not find compelling evidence to suggest that either of the metrics was significantly more important than the others.

\subsection{Feasibility of analysis}
Due to the implementation of the \sysname tool, the metric gathering stage has to launch multiple Amazon EC2 instances and run time-intensive metric scripts. In our experiments, the \sysname tool takes an average of about 400s to complete the analysis. Therefore, the analysis might be infeasible for small workflows with a small data source; however the approach is still valid for small workflows that are going to be run multiple times in the Cloud. Consequently, we suggest to use geographical distance as a crude indicator to initially rank Cloud regions and then to run the network latency and RTT analysis on the three top ranked Cloud regions from the previous ranking.

\section{Related work}
\label{sec:related}

This paper addresses the problem of \emph{where} geographically to deploy a workflow engine, given the specification of a workflow consisting of highly distributed services and a set of fixed points - in this case Amazon EC2 Cloud regions. As far as we know there has been no prior work on the topic of dynamically migrating workflow engines to Cloud-based resources in order to improve performance.


\subsection{Decentralised Orchestration} \label{section:dag_tools}


The concept of pointers in service-oriented architectures~\cite{soa_pointers} allows Web services to pass data by reference rather than by value. This has the advantage that the workflow orchestrator doesn't need to handle all data passing between the orchestrated Web services.

The \emph{Flow-based Infrastructure for Composing Autonomous Services} or FICAS~\cite{liu-data-flow} is a distributed data-flow architecture for composing software services.  Composition of the services in the FICAS architecture is specified using the Compositional Language for Autonomous Services (CLAS), which is essentially a sequential specification of the relationships among collaborating services. This CLAS program is then translated by the build-time environment into a control sequence that can be executed by the FICAS runtime environment.

\emph{Service Invocation Triggers}~\cite{binder} is an architecture for decentralised execution. Before execution can begin the input workflow must be deconstructed into sequential fragments, these fragments cannot contain loops and must be installed at a trigger.


In previous work~\cite{ccgrid08} \cite{hpdc08} we proposed \emph{Circulate}, a proxy-based architecture based on a centralised control flow, distributed data flow model. Our prior work~\cite{cite2} has also focused on decentralised service choreography models. In~\cite{IBM_decentral}, an architecture for \emph{decentralised orchestration} of composite Web services defined in BPEL is proposed.

All the approaches discussed in this section require either the workflow specification or the services involved in the workflow to be altered prior to enactment. In \emph{FICAS} the application code that is to be deployed needs to be wrapped with a FICAS interface; in the \emph{SOA pointers} and \emph{Triggers} approaches the workflow specification needs to be altered before enactment; the \emph{IBM} approach does not deal with the problem of where to geographically deploy an orchestration engine;  \emph{Circulate} like the other approaches require the addition of an extra actor, a proxy.  In contrast, our approach enables a workflow to be analysed and then dynamically migrated to a Cloud-based resource for execution, this avoids the costly setup cost of wrapping back-end services.



\subsection{Workflow Partitioning}

Workflow partitioning is an approach to divide a workflow into several sub-workflows, which are then executed on different sites. The most mature workflow partitioning mechanisms are contained in Pegasus \cite{paritioning}: a partitioner component decomposes an abstract workflow into smaller sub-workflows which are then mapped onto computational (usually Grid) resources.


Workflow partitioning can be a computationally ``expensive'' process and is primarily useful for very large-scale scientific workflows. We view workflow partitioning as a complementary activity; a potential area of further research is to use \sysname to schedule sub-workflows of larger-scale workflows.

\subsection{Third-party Data Transfers} \label{section:third}

This paper focuses primarily on optimising workflows where services are: not equipped to handle third-party transfers, owned and maintained by different organisations, and cannot be altered in anyway prior to enactment. For completeness it is important to discuss engines that support third-party transfers between nodes in task-based workflows.

\emph{Directed Acyclic Graph Manager (DAGMan)}~\cite{dagman} submits jobs represented as a DAG to a Condor pool of resources. Intermediate data are not transferred  via a workflow engine, instead they are passed directly from vertex to vertex. DAGMan removes the workflow bottleneck as data are transferred directly between vertices in a the DAG. \emph{Triana}~\cite{triana} is an open-source problem solving environment. It is designed to define, process, analyse, manage, execute and monitor workflows. Triana can distribute sections of a workflow to remote machines through a connected peer-to-peer network. \emph{OGSA-DAI}~\cite{ogsa-dai} is a middleware product that supports the exposure of data resources on to Grids. This middleware facilitates data streaming between local OGSA-DAI instances.


\subsection{Data-Aware and Location-Aware Scheduling} \label{section:dataaware}

Amazon have recently added Latency-Based Routing (LBR) \cite{lbr} to the Route 53 service. LBR provides functionally to reduce latency for end users by serving their requests from the region for which the network latency is lowest. LBR does not consider the complexities of where to deploy an application that is constructed from a number of highly distributed services.

Stork proposes an approach to data-aware scheduling \cite{data_aware_scheduling}: given an application dynamically deciding where to deploy the data. In contrast our approach decides where to deploy the application assuming that the data are fixed and cannot be relocated.

\section{Conclusion/Future Work}
This paper discussed how to increase the performance of highly distributed Web service workflows by dynamically deploying the workflow orchestrator on an IaaS Cloud rather than orchestrating remote services locally. We developed \sysname an analysis tool which, using the factors geographical distance, network latency and HTTP round-trip time, can analyse a given workflow and rank Amazon EC2 Cloud regions.

We ran several randomly generated workflows and found that orchestrating workflows in the Cloud significantly reduced execution time as well as the standard deviation of execution time. Overall, we concluded that geographical distance, network latency and HTTP RTT off the workflow correctly predict the best performing Cloud region to deploy the orchestrator.

Our proposed approach addresses the bottlenecks associated with executing highly distributed and data-intensive applications in the Cloud. The techniques discussed are general and can be applied to any workflow specification language and set of execution resources, e.g., we could easily add further IaaS nodes such as those provided by Rackspace.

Future work could potentially look at factors other than execution time. Using a Cloud cost forecasting system and different Cloud providers, the analysis could be extended to find the best Cloud region that minimises both total cost and execution time. Furthermore, issues such as the load of the web service nodes have not been considered in this paper and would make an interesting extension, as would adding throughput to the list of factors.





%

\bibliographystyle{abbrv} \bibliography{cloud}

\end{document}